\documentclass[12pt,preprint]{aastex}

\shorttitle{Probing the Solar Atmosphere}
\shortauthors{Penn, Schad \& Cox}

\begin{document}

\title{Probing the Solar Atmosphere 
Using Oscillations of Infrared CO Spectral Lines}

\author{M.J. Penn{\altaffilmark{1}},
T. Schad{\altaffilmark{1,2}},
E. Cox{\altaffilmark{1,3}} }
\altaffiltext{1}{National Solar Observatory{\altaffilmark{4}},
950 N Cherry Ave, Tucson, AZ 85718}
\altaffiltext{2}{Lunar and Planetary Lab,
University of Arizona, Tucson, AZ 85718}
\altaffiltext{3}{University of Arizona, Tucson, AZ 85718}

\altaffiltext{4}{NSO is operated by AURA, Inc.\
under contract to the National Science Foundation.}

\begin{abstract}
{Oscillations were observed
across the whole solar disk
using the Doppler shift and line depth of
spectral lines from the CO molecule
near 4666~nm
with the National Solar Observatory's
McMath/Pierce solar telescope.
Power, coherence, and phase spectra were examined,
and diagnostic diagrams reveal power ridges at the
solar global mode frequencies to show that
these oscillations are solar p-modes.
The phase was used to determine the height of
formation of the CO lines
by comparison with the IR continuum intensity phase shifts
as measured in 
\cite{kopp1992};
we find the CO line formation height varies from
$425 < z < 560$~km
as we move from disk center towards the solar limb
$1.0 > \mu > 0.5$.
The velocity power spectra show
that while the sum of the background and p-mode power
increases with height in the solar atmosphere
as seen in previous work,
the power in the p-modes only (background subtracted)
decreases with height, consistent with evanescent waves.
The CO line depth weakens
in regions of stronger magnetic fields,
as does the p-mode oscillation power.
Across most of the solar surface the phase shift is larger than
the expected value of
90~degrees
for an adiabatic atmosphere.
We fit the phase spectra at different disk positions with 
a simple atmospheric model 
to determine that the acoustic cutoff frequency is about 4.5~mHz
with only small variations,
but that the thermal relaxation frequency
drops significantly
from 2.7 to 0~mHz
at these heights in the solar atmosphere.}
\end{abstract}

\keywords{Sun: atmosphere, Sun: infrared, Sun: oscillations.}

\section{Introduction}

The solar infrared absorption lines
from the fundamental band
of the CO molecule
at 4666~nm
are well-known diagnostics
of the temperature minimum and chromosphere
of the solar atmosphere.
Initial discovery of off-limb emission in the quiet Sun from
these lines
\citep{noyes1972}
instigated the development of
thermally bifurcated models of the solar atmosphere,
\citep{wiedemann1994, ayres1996}
and the latest models draw particular attention to the dynamical events
and chemical equilibria in the atmosphere
\citep{ascensio2003, wedemeyer2005}.
The CO lines are known to show 5 minute period
Doppler oscillations
\citep{noyes1972, uiten1994, solan1996, uiten2000a}
which are presumably driven from the solar global p-modes,
and images of the solar surface in the lines
reveal that magnetic regions of the solar atmosphere show
a reduced molecular line strength
\citep{uiten2000a}.
Models of the line formation
\citep{uiten2000a}
suggest the
formation height of the strong CO lines at 4666~nm,
particularly the 3-2 R14 transition at 4665.8~nm,
varies from roughly 150 to 580~km above the
$\tau_{500} = 1$
formation height (z=0).

Studies of the solar p-modes at various heights in the solar atmosphere
have been done using a variety of spectral lines.
It is thought that the p-modes are reflected back into the Sun by
the steep gradient of density at the photosphere, 
and the wave solution requires an evanescent component
(characterized by an exponential amplitude drop and
a constant velocity phase with height)
to be present above the reflecting layer.
Simultaneous measurements in various spectral lines have shown
that for the solar p-mode frequencies the
velocity oscillations are in phase at different heights
\citep{lites1979}
but that the amplitude of the velocity oscillation increases with height
\citep{cram1977,ruizcobo1997}

In addition to velocity oscillations,
intensity oscillations are also
observed in the solar atmosphere;
since the early work of
\cite{schmeider1976}
the relationship between the two has become an important
diagnostic of the physical conditions of the solar atmosphere.
The phase between the temperature and velocity oscillations
at various heights in the solar atmosphere
has been probed by many authors
through the use of line depth and Doppler shift measurements 
of a different solar spectral lines
\citep{deubner1992, al1998, oliviero1999, strauss1999}.
A motivation here is to understand if the solar atmosphere reacts
adiabatically to the p-mode oscillations or not,
and if not, then to determine what other physical processes may be important.
Since the velocity oscillation amplitudes increase with height,
the solar atmosphere undergoes compression when the velocity
oscillations move the gas downwards towards the photosphere,
as the higher layers catch up to lower layers.
If energy is conserved by the atmosphere then
the gas temperature will reach a
maximum value when the compression reaches a maximum;
this corresponds to the time when the gas velocity is zero just before
it begins to move away from the solar photosphere.
If the gas temperature is plotted versus the atmospheric velocity
(defined as positive when the gas moves away from the photosphere,
which is the opposite direction as the observed Doppler shift)
in an adiabatic atmosphere
there will be a 90 degree phase shift between the two quantities,
with the temperature peak leading the atmospheric velocity peak.
The phase between the temperature and
the atmospheric velocity is referred to as the "I-V phase".
In summary,
observations from many different spectral lines show that
the I-V phase is less than 90 degrees in the photosphere,
around 90 degrees from the temperature minimum to the middle chromosphere,
and above 90 degrees in the high chromosphere
\citep{deubner1990, massiello1998}.

Complicating the spectral line measurements of I-V phase is the
fact that chromospheric spectral lines are often formed in a
non-LTE environment,
and the spectral line depth can change non-linearly
as the temperature of the solar atmosphere changes.
Investigations from Kopp and co-workers
\citep{kopp1990,kopp1992}
addressed this problem by observing temperature oscillations using 
thermal infrared continuum channels at wavelengths
from 50 to 800 microns from the airborne NASA
Kuiper Astronomical Observatory (KAO).
The continuum intensity depended only on the continuum opacity,
and so the observed intensities were linearly related to the 
temperature of the solar atmosphere
and provided excellent thermal probes.
These wavelengths sampled the solar atmosphere from heights
of roughly 340 to 820~km above z=0 and thus the temperature minimum
through low chromosphere could be explored.
Since the continuum observations lacked a way to determine the gas velocity,
simultaneous ground-based spectral observations were made using 
both photospheric and chromospheric spectral lines.

The KAO observations showed an I-V phase shift consistent with results
from spectral lines which sampled the same atmospheric heights.
In particular, the I-V phase was observed to be
122 degrees at 340~km,
and decreased with height to approach
95 degrees at 600~km;
the solar atmosphere became more adiabatic as one moved from the
temperature minimum up to the lower chromosphere.
In order to explain the non-adiabatic quality,
\cite{kopp1992}
introduced a simple thermal relaxation term into
the hydrostatic gas thermal balance equation
and solved for the phase in a simple model 
assuming an isothermal plane-parallel non-magnetic atmosphere.
The resulting analytic expressions were used to determine the
values of the thermal relaxation frequency $\omega_r$.

In the following paper we discuss our observations of temperature and
Doppler oscillations using the 4666~nm absorption lines of CO.
Since the CO spectral lines are formed in LTE conditions,
the line depth provides a more direct probe of the gas temperature
than other chromospheric lines;
and as spectral lines their Doppler shift gives
a direct measurement of the atmospheric velocity and thus they are
superior to observations using infrared continuum channels.
In Section~2 we discuss the data collection,
Section~3 we outline the data reduction and analysis procedures,
and in 
Section~4 we discuss the results obtained from the intensity, velocity,
and I-V phase measurements from these lines as they vary
across the solar surface and with height in the solar atmosphere.
We apply the simple analytic model of
Kopp
to our I-V phase spectra
to estimate the gas properties in the solar atmosphere.

\section{Observations}

The observations presented here were collected on 16 September 2009 
at the National Solar Observatory McMath-Pierce solar telescope
\citep{pierce1964}
using the 13.5~m main spectrograph.
In order to image the full solar disk onto the detector,
the "integrated light" flat mirror
was inserted into the normal optical path of the telescope,
thus providing unfocused sunlight into the observing room.
The solar image was formed with a
70~mm diameter 1400~mm focal length
(effective focal length at 4666~nm)
CaF2 singlet lens, 
which formed a 13.2~mm diameter prime focus image on the slit
of the main spectrograph.
The resulting f/20 beam overfills the spectrograph collimating mirror,
reducing the transmitted solar flux and
increasing the spectrograph stray light,
since the spectrograph is designed to work at f/60.
These sacrifices were made in order to quickly scan the entire solar disk.

At the spectrograph exit port,
the NSO Array Camera (NAC) was fed with a single fold mirror.
The NAC is a cryogenically cooled
Aladdin III InSb 1024x1024 pixel array detector.
The beam is fed through a cold (55K) order sorting filter
to a single cold fold mirror and then to the InSb array;
for these observations a 25~mm diameter filter
centered at 4666~nm with a band-pass of 17~nm was used.
The NAC was operated in "streaming" mode
where about
4.5 frames per second were
collected and stored,
during which time the solar image was scanned back and forth across
the spectrograph slit asynchronously
by moving the McM-P heliostat mirror (M1).
The scanning sequence took about 50 seconds to cross the solar disk,
and 160 scans of the solar disk were taken during 133 minutes
from 1647 to 1900 UT.

The CaF2 lens determines the spatial sampling along the spectrograph slit,
and here we obtain 3.92 arcsec per pixel.
The scan rate and camera readout time determine the spatial sampling
in the direction perpendicular to the spectrograph slit, and
there we achieve
12.39 arcsec per pixel.
At this wavelength the spatial resolution of the 70~mm diameter
lens is
13.75 arcsec
and so the image is oversampled along the slit but not
Nyquist sampled in the scan direction.
The dispersion of the spectrograph is
0.0067~nm per pixel,
but the spectral resolution is estimated
to be ten times less at
$R = 60,000$
by comparing the observed spectra with
model spectra provided by Uitenbroek.
Simulations degrading the model spectra
to this resolution also show that
there seems to be significant scattered light,
likely caused by feeding the spectrograph with
a faster beam than its design and the resulting
internal reflections.

\section{Data Reduction and Analysis}

The NAC was used in single-frame read streaming mode
instead of using double-correlated sampling;
in this way the InSb array pixels
were read-out just once per exposure
and the cadence was fast.
NAC exposure times were very short,
only 50~milliseconds per frame.
Tests of this mode have determined that the entire array bias level
changes slightly from frame to frame,
but that individual pixel biases remain constant.
Thus in addition to the usual dark current subtraction procedure,
a set of pixels which were not exposed to light from the spectrograph
were used to remove the changing bias level.
A list of bad pixels
(less than 300 of the one million pixels)
was computed from a dark frame and the values of
these pixels were replaced with a linear extrapolation
from neighboring pixels.
Gain correction was simply accomplished by scanning the spectrograph
grating while taking a sequence of images,
and selecting the maximum
intensity value for each pixel out of the resulting data cube.
This intensity level occurs when a pixel is exposed to the local continuum,
and in this way the spectral lines are removed from the gain correction
image.
The slowly changing transmission profile of the cold order-sorting
filter remains in this procedure,
but it is removed with a linear fit to the intensity of the
pixels which observe the continuum spectrum.
Figure~1 shows a sample spectral frame with several spectral lines
labeled.
The signal-to-noise ratio of the data can be estimated with the
standard deviation of the continuum pixel intensities to
be roughly 50.

Three strong CO spectral lines were then fit in the corrected data frames.
Routines were used to
fit both Gaussian
(from {\it IDL})
and
Voigt functions
(written in {\it IDL} by D.M. Zarro)
to the line profiles.
Initial guess for the position of each spectral line accounted for
the spectrograph drift seen in the data
(roughly 0.01~nm per hour)
and the Doppler shift from solar rotation;
in this way neighboring solar or atmospheric spectral lines were avoided.
As part of the Voigt fitting, the parameters from the Gaussian fit were
used as a first guess.
Because of this the Voigt fitting procedure was more sensitive to noise.
A comparison of the line center wavelengths returned from the Gaussian and
the Voigt fitting was done;
no systematic differences were seen between the values, 
and the small differences between the line centers were consistent with
expectations from spectral measurements with a signal to noise of about 50.
Since the simpler Gaussian fitting was able to fit more spectral lines
successfully and no differences were seen between the two methods
when both fits converged,
the Gaussian fit values were used in the rest of the analysis.

Scanning with the heliostat mirror (M1) at the McM-P without
any limb guiding introduced significant image drift
(about 100 arcsec E/W and 20 arcsec N/S)
during the 2 hour observing run.
Because the image scanning extended above the solar limbs,
the image drift was accommodated within the field-of-view of the observation.
The solar images were registered using the continuum solar disk.
Because the massive 2~m diameter M1 has a limited scan speed, 
the image scanning proceeded East to West in one scan, and then
back West to East in the following scan,
rather than having a "flyback" to scan continuously in one direction.
In this way pixels on the solar disk are sampled at different times.
To correct for this, alternate pairs of scanned images were linearly 
interpolated to form a new image at a single mean time.
Finally, irregularities and dust on the spectrograph slit produced a
streaking in the data, and this was removed.
A co-registered and calibrated time series of the spectral line depth,
the line position and line width was produced for each spectral line.
In order to improve the signal to noise ratio of these parameters,
the quantities from all three strong CO lines were averaged together.
While the lines may sample
slightly different regions of the solar atmosphere,
since the depths of the three lines vary by less than 3\%
from each other
\citep{wallace2003},
we assume that they originate from roughly the same heights.

Following
\cite{kopp1992}
a set of 10 pixels at disk center were averaged together,
and the mean atmospheric velocity and line depth
were plotted as a function of time.
A cross-correlation function was computed
between the velocity and line depth;
the peak of the cross-correlation function revealed that the
temperature led the atmospheric velocity signal on average by
a little more than 90 degrees (assuming a dominant 
oscillation period of 5 minutes).
This cross-correlation procedure was then repeated for each pixel
across the solar surface, and a map of the resulting I-V phase was produced.

A more detailed analysis of the I-V phase was produced with a 
Fourier decomposition of the velocity and line depth time series
this analysis follows standard techniques used by
\cite{lites1993} and
\cite{krijger2001}.
The individual time series were transformed to produce maps of the
power in the atmospheric velocity and line depth parameters
across the solar disk.
If
$f_v(t)$
is the time series of the atmospheric velocity,
we define
$F_v(\nu)$
as the Fourier transform of the atmospheric velocity.
As discussed in
\cite{krijger2001},
we can decompose the power spectrum into real and
imaginary components
$F_v(\nu)=a_v(\nu)+ib_v(\nu)$.
We define similar quantities for the line depth as
$f_d(t)$
and
$F_d(\nu)$.
The cross-power spectrum was computed using
$CP_{dv}(\nu)=F_d(\nu)F^{*}_v(\nu)$,
where
$F^{*}_v(\nu)$
is the complex conjugate of the transform of the velocity time series.

From these we now compute the coherence and 
the phase spectra in the normal way,
except that the particular way of computing the ensemble averages
deserves mention.
We can express the cross-power spectrum in terms of
real and imaginary components as
$CP_{dv}(\nu)=c_{dv}(\nu) + i d_{dv}(\nu)$.
We use these complex coefficients for the ensemble averaging.
In particular, 
for the center-to-limb phase analysis,
we average the
$c_{dv}(\nu)$
and 
$d_{dv}(\nu)$
coefficient in each bin of pixels at a given value of
$\mu$,
and then compute the phase spectrum using
$\phi(\nu)=atan(\frac {d_{dv}(\nu)}{c_{dv}(\nu)})$.
For the diagnostic diagrams we average the
complex coefficients over
$k_x$
and
$k_y$
(the wavenumbers in the N/S and E/W solar directions)
during the computation of 
the average horizontal wavenumber
$k_h = \sqrt{k_x^2+k_y^2}$.
As pointed out in
\cite{krijger2001},
averaging the complex coefficients
before computing the 
phase
reduces the noise in the calculation,
whereas computing the phase first and averaging the noisy
phase values drives the computed phase to zero.

One aim of this study was to look for center-to-limb variations
in several oscillation parameters,
and then to express these as a function
of height through the solar atmosphere.
Some studies of oscillations have been done at different limb
positions before
\citep{solan1996, uiten1994}
however having the entire disk available in this data set
provided a considerable advantage.
Since the data contain about 60,000 pixels across the solar disk,
binning in annuli of
$\Delta\mu=0.01$
enabled us to average our oscillation parameters using
between 600 and 1200 pixels.
This substantially increases the signal-to-noise ratio
in our data
when compared with a single measurement at a given value of
$\mu$.

Finally the two-dimensional aspect of this data enabled the
production of diagnostic diagrams
(or l-$\nu$ diagrams).
Many studies have been done to examine the I-V phase using
diagnostic diagrams with other spectral lines
\citep{al1998, oliviero1999, strauss1999, deubner1992}
but our data set sampled
a unique atmospheric height over a unique range of spatial wavelengths.
For this analysis,
a section of roughly 
1000x1000 arcseconds from the disk center was extracted
from both the velocity and line depth time series.
No correction was made for projection effects or 
differential solar rotation,
nevertheless valuable information regarding the spatial and spectral
dispersion of the observed oscillations is obtained from this analysis.

\section{Discussion}

The shifts of the CO line centers across the solar disk are consistent
with the expected solar photospheric rotation velocity.
After removing solar rotation,
a low pass temporal filter was applied to the time series in order to
extract the supergranulation component of the disk velocity.
In these images the supergranulation cells are clearly visible,
and a fit to the outflow velocities 
(both positive and negative lobes)
suggests a horizontal outflow velocity of
186~m~s$^{-1}$.
This is in agreement with the distribution of outflow 
velocities seen from Hinode feature tracking measurements
in the UV continuum channels (Tian et al 2010)
which originate from the same approximate heights.
The behavior of the supergranulation outflow Doppler signals
with $\mu$ suggests a height variation in the outflow speed,
which will be investigated in future work.
An analysis of the line bisectors of the CO lines showed no variations,
which was likely due to the rather low spectral resolution achieved
with these observations.

\subsection{Spatial Variations}

Using the low pass temporal filter, maps of the line depth
across the solar disk were made.
The data show that on a global scale
the CO line absorption appears weakened in magnetic
features associated with the solar magnetic network;
confirming results from previous higher resolution studies
\citep{uiten1994, ayres1996}.
In Figure~2 we show an image produced from a magnetogram
taken on the same day from the NSO SOLIS instrument
(i.e. \cite{jones2002});
here the absolute value of the magnetogram is taken, 
and then the map is smeared with box-car averaging to a 
resolution of about 10 arcseconds
(a little sharper than the expected CO map resolution).
In Figure~2 we plot the mean CO line depth
(averaged over the three spectral lines)
across the solar disk,
excluding pixels near the limb where seeing and fitting noise
dominate the map.
There is a very clear correspondence between the position of 
stronger magnetic fields in the SOLIS map with the regions of weaker
CO line depth.

The beating of solar p-modes with slightly different frequencies
complicate the simple cross-correlation of the line depth and velocity
time series.
This is especially true if the power distribution among the oscillation
modes varies in the line depth compared to the velocity measurements,
and such a different power distribution is a natural consequence of 
non-adiabatic atmospheric conditions.
Furthermore, if the I-V phase varies with oscillation frequency we
expect even more difficulty with a simple cross-correlation technique.
Nevertheless computing the cross-correlation was very 
robust.
As previously noted,
the cross-correlation lag was measured in
units of seconds of time,
but then was converted to an angular phase shift
assuming a dominant oscillation period of five minutes.
In most cases, the peak of the cross-correlation function occurred with the
line depth leading the velocity by about 90 degrees;
in some cases however
the peak in the cross-correlation function at -270 degrees was higher.
To unwrap the measurements,
we forced our analysis software to fit the
peak near 90 degrees.
Figure~2 shows a map of the phase shift across much of the solar disk.
As the Sun had very low activity, only small variations are noted;
however these variations are significant, and are not tightly correlated
with the line depth map of Figure~2b.
The phase map shows something different than the line depth map, 
and presumably the variations represent regions of more or less adiabatic
response in the solar atmosphere.
As we will discuss in detail later, one interpretation of this is that
the phase map show regions with faster and slower thermal relaxation.

The Fourier analysis of the line depth and velocity time series 
was done to more fully investigate the properties of the oscillations.
Both line depth and velocity showed power spectrum peaks at the 
roughly 3.3~mHz, equivalent to a period of five minutes.
Unlike other studies of oscillations with these spectral lines,
these data show no power for oscillations near 
5~mHz
(3 minute period).
It must be stated that the signal-to-noise ratio of these
data benefit significantly from the ability to
average across many pixels from the solar surface;
in some of the following discussions we average over tens of
thousands of pixels,
and so our power spectra has a factor of 100 times better
signal-to-noise than previous works which report
spectra from single positions on the solar disk.

The oscillation power integrated from
$ 3.0 < \nu < 4.2$~mHz
using the CO velocity power spectrum
measured pixel-by-pixel
across the solar disk
is also shown in Figure~2.
Regions of high power and lower power are clearly seen.
By examining the
500 weakest and strongest line depth pixels on the 
CO line depth map,
(also corresponding to regions of stronger and weaker line-of-sight
magnetic fields)
a power deficit in magnetic regions is seen as the p-mode power
(after background subtraction).
In the regions of weaker line depth pixels
(stronger magnetic fields)
the power is about a factor of 0.5 times
the power in the strongest line depth pixels.
There is also a weak correspondence between
oscillation power and phase,
with regions of
high (low) power
showing
non-adiabatic (adiabatic) phases.

The interest in observing the full solar disk with these
observations was to look for both large and small spatial scale
variations of the oscillations parameters.
The variation from center-to-limb can be used to characterize
the height variation within the solar atmosphere
by using the Eddington-Barbier relationship.
Here the approximation is made that the source function
arises solely from a thin layer at
$\tau=1$,
and that the height of this layer changes
across the solar surface with
$\mu$.
Using our measured I-V phases, it becomes
possible to convert from
$\tau$
to height
$z$
in the solar atmosphere.
As discussed by Kopp et al. (1992),
the formation of the infrared continuum in the solar atmosphere
is more easily understood than the formation of spectral lines,
as the continuum is formed in LTE and the dominant source
of opacity is H$^{-}$ free-free absorption.
Given a solar atmospheric model,
(they use model~C from
\cite{vernazza1981})
they determine the formation heights of their
50, 100, 200, and 400~$\mu$
IR continuum channels as
340, 420, 480, and 600~km
above
$z=0$.
By subtracting their observed phase shifts
from 180 degrees
(since they use Doppler velocity rather than atmospheric velocity),
we can compute I-V phase lags for these heights as
122, 122, 109 and 95 degrees respectively
(with errors of about $\pm$10 degrees).

Figure~3 shows the mean I-V phase shift as a function of
$\mu$
from our CO observations.
Here we have averaged the phase spectra from
3.0 to 4.0~mHz,
and averaged over annuli on the solar disk
with a width of
$\Delta\mu = 0.01$.
The I-V phase drops from about 116 degrees at disk center
to about 100 degrees at 
$\mu=0.5$.
These values are clearly within the range of phases 
observed by Kopp et al. (1992)
and we can now determine the atmospheric heights
from which they arise,
assuming that equal phases in our data correspond to equal heights
in the IR continuum data.
In detail,
we use the Eddington-Barbier relation to 
replace
$ln(\tau)$
with
$ln(\mu)$.
We determine the relationship between
$z$
and
$\phi_{I-V}$
from the data from
\cite{kopp1990},
and then use the relationship between
$\mu$
and
$\phi_{I-V}$
in our observations to determine a
$ln(\mu)$
vs
$z$
function for the formation height of our CO line depth.
Our CO I-V phases between 
$0.5 < \mu < 1.0$
correspond to heights between
$560 > z > 425$~km.
and the scale height which we derive for the variation of the 
CO I-V phase gives
$H_1=195$~km.

We can now examine the power in the CO oscillations
as a function of disk position, and therefore height.
Figure~4 shows the CO velocity power spectra averaged
over the spatial positions from
$0.99 < \mu < 1.00$.
It shows a significant background
power which varies with frequency,
and a broad peak near 3.3~mHz which represents the well-known
p-mode oscillations.
We fit a polynomial function to the log of this background using the
frequency ranges of
$1.5 < \nu < 2.0$~mHz
and
$5.0 < \nu < 8.0$~mHz
and interpolate the background power through the other frequencies.
After subtracting this background,
we fit a Gaussian function to the power spectrum.
The fits to the power spectrum are shown as the solid line in 
Figure~4.
The fits to the oscillation power across
the solar disk were examined.
The peak of the p-mode power given by the Gaussian
fit was corrected for two projection effects:
first, the power was divided by
$\mu$
to compute the oscillation power assuming purely radial motions,
and second
the power was corrected for the smearing in spatial wavenumber
caused by projection effects as one moves from the center to the limb.
(This correction,
basically a distribution of power with spatial wavenumber,
was computed by averaging the velocity signal in
pixels near disk center and then computing the power spectra of
these averaged velocities.)
After making these two correction,
we transform the
$\mu$
coordinate into a height coordinate
$z$
and examine the power changes with height.

Figure~4 shows that
the total power (p-mode and background) increases
with height,
whereas the p-mode power alone
decreases with height.
The increase of the total power is consistent with
previous studies
\citep{ruizcobo1997, simoniello2008}
and supports the idea that the velocity oscillation power must
increase with height as the density drops.
An exponential fit to this background power gives a scale height of
$H_2=230$~km,
consistent with values used for chromospheric models.
By examining the p-mode power alone, after subtracting the background power,
we find that it
decreases with height, as would be expected from
an evanescent solution to the wave equation.
This decrease is very rapid;
the scale height is much smaller and consistent
with photospheric scale heights,
$H_3=90$~km.

Another parameter derived from the power spectra fitting is the
peak central frequency for the p-mode oscillations.
While many studies find 3-minute period oscillations at 
chromospheric heights and lower
(using these CO spectral lines)
we find no significant power peaks at that frequency.
Moreover, the central frequency position of the fit to the p-mode
power does not change with height through the range of
$560 > z > 425$~km.
No 3-minute period power is seen in
an analysis of the line depth oscillation power spectra,
and the central frequency of the p-mode oscillations
as measured with the line depth
also remains constant with disk position. 

\subsection{Diagnostic Diagrams}

In Figure~5 we present diagnostic diagrams,
or l-$\nu$ diagrams
of several parameters from the CO data;
these diagrams are produced using the three dimensional
Fourier transform of the region near the center of the solar disk
as discussed in
Section~3.
A simple correction for solar rotation was made by aligning
the ridge structure for positive and negative
frequencies before coadding;
but no correction is made for projection effects.
The CO velocity 
l-$\nu$ diagram
clearly shows several ridges of power, 
and these six ridges are at the position of the well-known
n=2 through n=7 
ridges for the 
global oscillation modes.
The CO line depth diagram is noisier,
but shows a couple of ridges at the positions of the
global oscillation modes.
In both parameters, 
no distinct ridge structure is seen 
at high frequencies.
At 
$4< \nu < 5$mHz
the ridges fade into the background power;
and no other structure is seen at the 
$\nu=5$mHz
frequency where some studies
have reported 
to have seen chromospheric oscillation power.

The coherence between the velocity and line depth
in Figure~5
also shows strong ridge structure.
While the resolutions
$\Delta$l
and 
$\Delta\nu$
not especially high,
there is some evidence that the coherence between the 
global p-mode ridges does not go to zero.
This is consistent with 
\cite{oliviero1999}
who show significant background power at the
inter-ridge positions in
measurements of global p-modes
near these spatial frequencies.
The coherence also shows
significant non-zero values at
high temporal frequencies,
up to about 
$\nu=6.5$mHz,
especially at low spatial degree.
Although there are no ridges or peaks at these frequencies
suggesting distinct oscillation modes,
this large coherence value suggests
that waves at this chromospheric oscillation 
frequency may be seen with these CO lines.

The I-V phase diagnostic diagram
in Figure~5
shows a correlation with the
coherence diagram: 
the phase values are noisy where the coherence is small, 
and the phase values change smoothly where the coherence is large.
There is very little evidence for a ridge structure in this data,
and the phase is rather constant at all spatial wavelengths.
The phase does show a gradual decrease as one moves 
to lower temporal frequencies,
from about
$\phi=130$~degrees at
$\nu=2.5$~mHz
to roughly
$\phi=90$~degrees at
$\nu=4.5$~mHz.
(It should be noted that trend this agrees with
$180 -\phi$ from
\cite{oliviero1999},
assuming that the phase in that work was measured using
Doppler velocities.)
There is no evidence for a sharp change in the phase
or a plateau of low phase
below
$\nu=2$~mHz
as seen in other work
\citep{deubner1990};
as suggested by the low coherence at these frequencies,
this may simply reflect that the phase in these
observations is not well measured here.

\subsection{Phase Spectra Fits with Analytic Model}

The Kopp model
\citep{kopp1990, kopp1992}
is based upon several simplifying assumptions.
First it uses hydrostatics,
thus ignoring the gas dynamics
and the magnetic fields which are present in the solar 
temperature minimum and chromosphere
(effectively the magnetic field
is assumed to be less than about 200 Gauss).
The model is based upon an isothermal atmosphere which simplifies the
temperature profile in this region of the atmosphere.
And finally, the model assumes a plane-parallel atmosphere which may
have implications with these full-disk measurements.
Nevertheless,
the Kopp model is one of the few analytic models
(in addition to 
\cite{worrall2002})
to date which
describes the I-V phase for solar oscillations in physical parameters;
because of this it is still very useful.

\cite{kopp1990}
develop expressions for the I-V phase in terms of the
atmospheric acoustic cutoff frequency
$\omega_c$,
the scale height
$H$
the ratio of specific heats
$\gamma$
and the thermal relaxation frequency
$\omega_r$.
The work makes no assumption regarding the actual physical mechanism
for the thermal relaxation process,
it just assumes that the process follows a Newton cooling law.
What is explicitly lacking from the analysis of
\cite{kopp1990}
is a direct expression for the I-V phase spectrum,
$\phi(\nu)$.
In the Appendix below, we continue the
development from 
\cite{kopp1990}
one extra step to derive and expression for
$\phi(\nu)$
which depends solely on the physical variables listed above.
Our expression is listed in Equation 5 of the Appendix.

We use our expression for
$\phi(\nu)$
to fit our I-V phase spectra as determined at different disk positions
from our data.
The points Figure~6 shows a sample I-V spectrum measured near disk center.
An adiabatic model with no thermal relaxation would produce a flat
spectrum with a 90 degree phase shift at all frequencies;
this is clearly contradicted by the data which show a decrease from
about 120 degrees to about 105 degrees from frequencies of
about 2.8 to 4.0 mHz
(these frequencies are where the I-V coherence is high).
Following 
\cite{kopp1990}
we can assume that the acoustic cutoff frequency is 
$\omega_c / (2\pi) =4.5$mHz
and do a one parameter fit to the data to determine the best value of
the thermal relaxation frequency
$\omega_r$.
This 1d fit is shown as the dashed line in Figure~6;
it has a high
$\chi^2$
value and does not produce a satisfying fit to the data points.
If we let the acoustic cutoff frequency be a free parameter and
do a 2d fit to the spectrum,
we significantly reduce the value of 
$\chi^2$
and derive a more satisfying fit to the data points.

An examination of the 
$\chi^2$
surface which results from a 2d fit using 
$\omega_c$
and
$\omega_r$
reveals that in some cases the analytic function only weakly constrains
$\omega_c$.
This combined with the inherent non-linearity of the
$atan$
function have foiled our attempts to fit this data using standard
non-linear least squares software.
Instead,
for each I-V spectra at a given disk position
($\mu$)
we simply examine the 
$\chi^2$
over the range of
$0 < \omega_r /(2\pi) <10$mHz
and
$0 < \omega_c /(2\pi) <10$mHz
and find the point of minimum
$\chi^2$.
We estimate the 68\% confidence ranges
by using specific
$\Delta\chi^2$
intervals on this surface according to standard numerical analysis techniques
\citep{press1994}.

Figure~6 shows the values we derive from fitting the I-V phase
spectra over the range of
$0.25 < \mu <1.0$mHz.
As in the case of Figure~6,
we plot the results for the 1d fit using the dashed line,
and the 2d fit results using solid lines.
In the 1d case we see that 
the thermal relaxation frequency
$\omega_r$
drops from the center of the disk to the
limb;
this suggests that as the data probe higher layers of the solar atmosphere
the thermal relaxation process becomes less important and 
the atmosphere becomes more adiabatic.
The same behavior for
$\omega_r$
is shown with the 2d fits
where the relaxation frequency decreases at 
$\mu$
decreases,
although the values are consistently 0.5mHz higher
than in the 1d fits.
And finally, 
the value for the acoustic cutoff frequency is around 4.5mHz,
although the data suggest the value may significantly drop
near the solar limb
(and thus at higher heights in the solar atmosphere).

The values of
$0.5 < \omega_r /(2\pi) < 2.8$mHz
correspond to a cooling time scales between
2000 and 360 seconds
which are longer than
the values of
290 to 76 seconds
for this range of heights in the solar atmosphere
measured with different fitting methods in 
\cite{kopp1992}.
While the relaxation rate
(defined by 
\cite{kopp1990})
of
${\omega_r} \over {\omega_c}$
which we obtain is similar to
the values determined by 
\cite{kopp1990},
our values of 
$\omega_r$
are somewhat lower since our fitted values of 
$\omega_c$
are somewhat lower than the 4.5mHz assumed in their work.

\acknowledgments

We would like to acknowledge Claude Plymate and Eric Galayda
for exceptional support
during the observations at the McM/P telescope,
and Han Uitenbroek for the model CO spectra
used to determine the spectral resolution of
the observations.
E. Cox was supported through the National Solar Observatory 
Research Experiences for Undergraduate (REU) site program,
which is co-funded by the Department of Defense
in partnership with the National Science Foundation REU Program.

{\it Facilities:} {McMath-Pierce}.

\pagebreak

\pagebreak

\section{Appendix: Analytic Expression for the Phase Spectrum
Including Thermal Relaxation}

Kopp (1990) develops an analytic expression for the dispersion relation
for a thermally relaxing chromosphere.
Assumptions in his simple model include
an isothermal atmosphere near the temperature minimum,
no magnetic fields,
the plane-parallel approximation,
and a Newton's cooling law for the thermal relaxation process.
While these assumptions may be a poor reflection of the actual
physics of the solar chromosphere
the work gives one of the few clear analytic expressions
for the I-V phase relation
published to date.
We continue the discussion from Chapter~5 of Kopp (1990) a few
steps to determine an expression for the I-V phase as a function
of frequency from his model.

Kopp starts with
the conservation of mass,
the equation of motion,
a thermal equilibrium equation including Newton cooling,
and the ideal gas law
(see his eq. 5.4 through 5.7).
He linearizes the equations around an equilibrium solution,
assumes vertical perturbations in pressure, density and velocity,
(where the vertical velocity perturbations are
$ \vec{v} = w \vec{z} \exp^{i(\omega x - k_z z)} $
with 
$w$
as a unitless constant)
and determines the dispersion relation for vertical wavenumber 
$k_z$
as a function of oscillation frequency
$\omega$
as follows (Kopp's equation 5.25):
\begin{equation}
ik_z = \frac{\omega_c}{c_s} \left( -1 \pm \sqrt {1. - \frac{1-i(\omega_r/\omega)}{1-i(\omega_r/\gamma\omega)}\left( \frac{\omega}{\omega_c} \right)^2} \right)
\end{equation}
where 
$\omega_c = \frac{c_s}{2H}$
is the acoustic cutoff frequency,
$c_s$
is the sound speed, 
$\gamma$
is the ratio of heat capacities,
$H$
is the scale height and 
$\omega_r$
is the thermal relaxation frequency.

By assuming that
$k_z$
can be written in the form
\begin{equation}
k_z = k_r + i k_i
\end{equation}
where
$k_r$
and
$k_i$
are purely real values,
and by writing an expression for temperature perturbations in a similar
form to the velocity perturbations
$ \delta T = \xi T_0 \exp^{i(\omega x - k_z z)} $
with 
$\xi$
as a unitless constant,
Kopp relates the temperature and velocity perturbations as
(equation 5-29):
\begin{equation}
\xi = \left[ \left( \frac{\gamma \omega k_r / c_s^2} {k_r^2 + (k_i - 1/H)^2} - \frac{k_r}{\omega} \right) -i \left( \frac{(\gamma \omega / c_s^2)(k_i - 1/H)}{k_r^2+(k_i-1/H)^2}+\frac{k_i}{\omega} \right) \right] w
\end{equation}
Finally, the I-V phase can be determined by (equation 5.30):
\begin{equation}
\phi = \arctan{ \left( \frac{Im(\xi/w)}{Re(\xi/w)} \right) }
\end{equation}

To develop an explicit expression for the phase as a function of frequency,
we define the ratios
$ a \equiv \omega / \omega_c $
and
$ b \equiv \omega_r / \omega_c $
and gather terms in
Equation~1.
Next we reduce the square-root of the complex number and solve
for
$k_r$
and 
$k_i$,
keeping only real solutions for these two coefficients.
Finally we substitute the wavenumbers into the phase expression to derive:
\begin{equation}
\phi = \arctan { \left[ \frac {\frac{-\gamma}{(2H)^2}a^2(k_i-1/H) - k_i(k_r^2 + (k_i-1/H)^2)} {\frac{\gamma}{(2H)^2}k_r a^2 - k_r(k_r^2+(k_i-1/H)^2)} \right]}
\end{equation}

where:
\[ k_i= \frac{1}{2H}(1-f) \]
\[ k_r= \frac{1}{2H} \frac{1}{\sqrt{x_1}} \frac{x_3}{2f} \]
\[ f = \frac{1}{\sqrt{2x_1}}\sqrt{x_2+\sqrt{x_2^2+x_3^2}} \]
\[ x_1 = a^2\gamma^2+b^2 \]
\[ x_2 = a^2\gamma^2(1-a^2)+b^2(1-a^2\gamma) \]
and
\[ x_3= a^3b\gamma(\gamma-1) \]
As a check for this solution, we reproduce Figure~5.2 of Kopp (1990) using
values for 
$\gamma = 5/3$
and
$H = 150$~km.
We also select the solution for 'upwardly localized waves' (not shown here)
and successfully reproduce Kopp's Figure~5.3.
Section~4.3 of this paper discusses the agreement
between our Equation~5 and our observations
and the detailed fitting we do using this expression to derive
the behavior of 
$\omega_c$
and
$\omega_r$
in the solar atmosphere.

\clearpage
\begin{figure}
\plotone{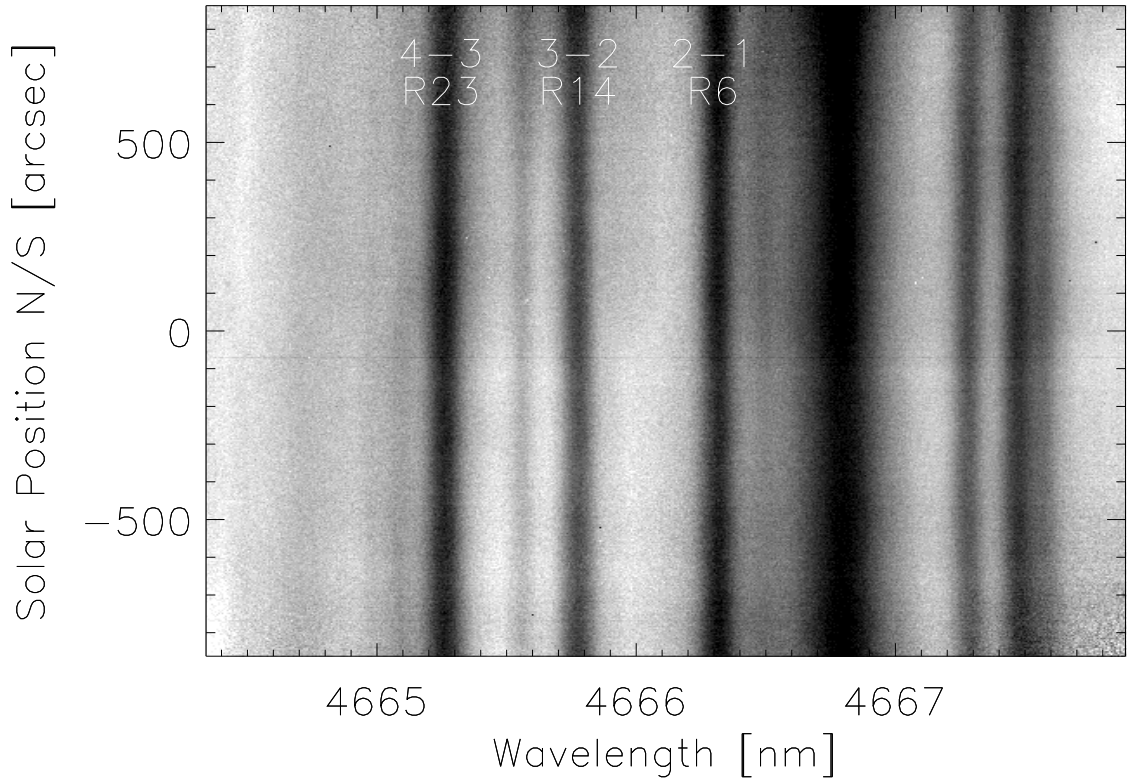}
\caption{A spectral frame from
the solar disk scan at 1647UT on 16 Sep 2009 using the NAC.
The three CO lines used in this study are labeled;
the other features in the spectrum
are all CO lines except for the 
broad atmospheric absorption at 4666.8nm.
(Identifications are from 
\cite{wallace2003}).
The spectrograph slit roughly lies
along the central meridian of the solar disk in this frame.
}
\end{figure}

\clearpage
\begin{figure}
\plotone{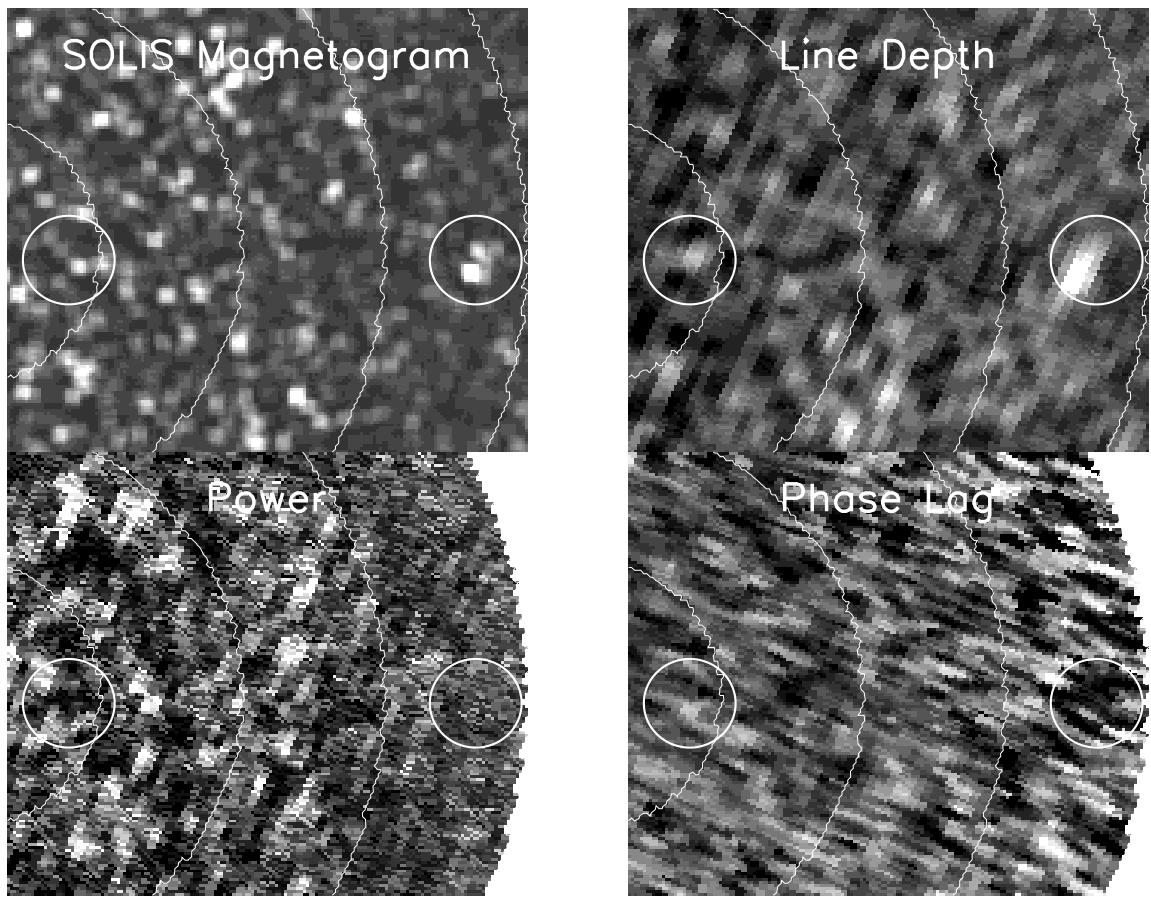}
\caption{Maps of
the absolute values of the SOLIS magnetogram
(smeared to lower spatial resolution),
the CO line depth,
the CO Doppler velocity p-mode power,
and the CO I-V phase compute from the cross-correlation of the
line depth and atmospheric velocity time series.
Solar north is up and solar west is to the right in each of the images.
Each image has circles at 0.2, 0.4, 0.6 and 0.8 solar radii, and
the center of the disk is near the left edge of each image.
(The power and phase maps are only displayed up to 0.8 solar radii.)
The two smaller circles show regions of reduced CO line depth,
which correspond to regions of vertical magnetic field in
the SOLIS magnetogram.
Both regions show reduced p-mode power,
but while the phase lag in these regions is weakly 
correlated (showing phases close to 90 degrees)
it also shows uncorrelated structure.
The phase map is scaled from about 130 degrees (white) to 90 degrees (black).
}
\end{figure}

\clearpage
\begin{figure}
\plotone{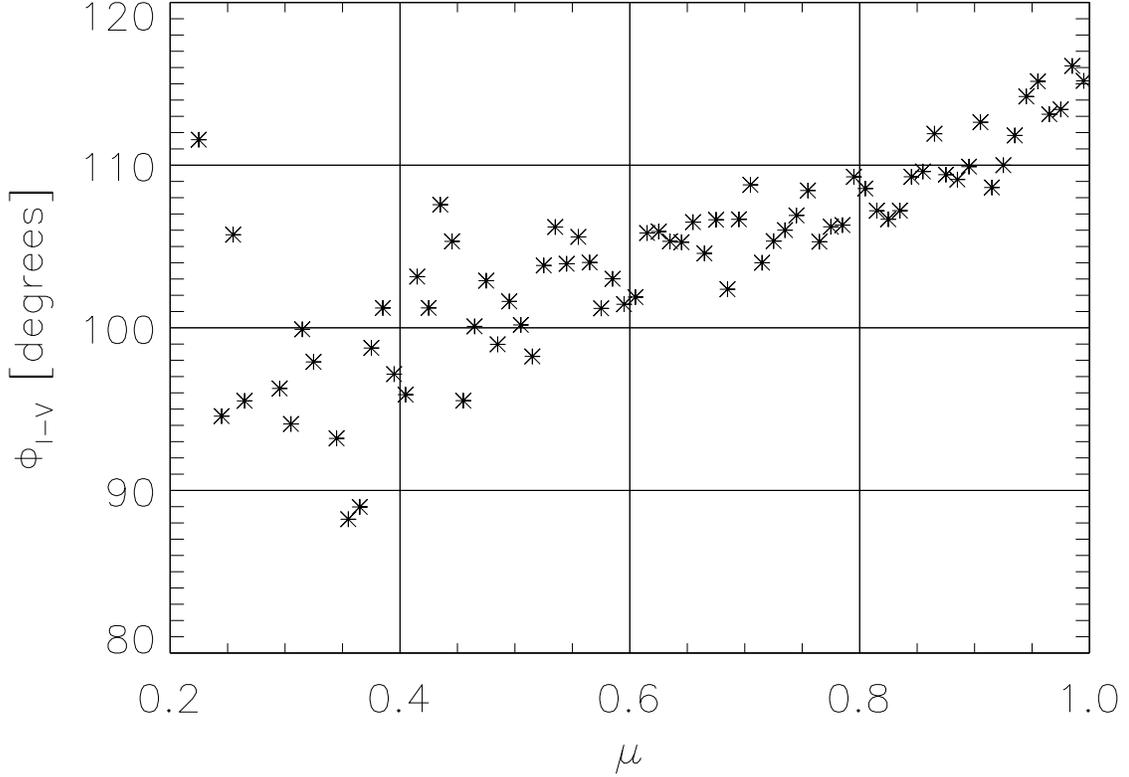}
\caption{The I-V phase shift averaged from
$3.0 < \nu < 3.4$~mHz
as a function of position across the solar disk,
from the center of the disk
$\mu=1$
toward the limb.
The value of
$\phi_{I-V}$
would be equal to 90 degrees if the solar atmosphere responded
adiabatically,
but this is clearly not the case over much of the solar surface.
Using the Eddington-Barbier relationship,
we can connect a decreasing value of 
$\mu$
with larger heights in the solar atmosphere;
and so the trend of
$\phi_{I-V}$
to move towards 90 degrees as 
$\mu$
decreases means that the atmosphere behaves more adiabatically with 
increasing height.
}
\end{figure}

\clearpage
\begin{figure}
\plottwo{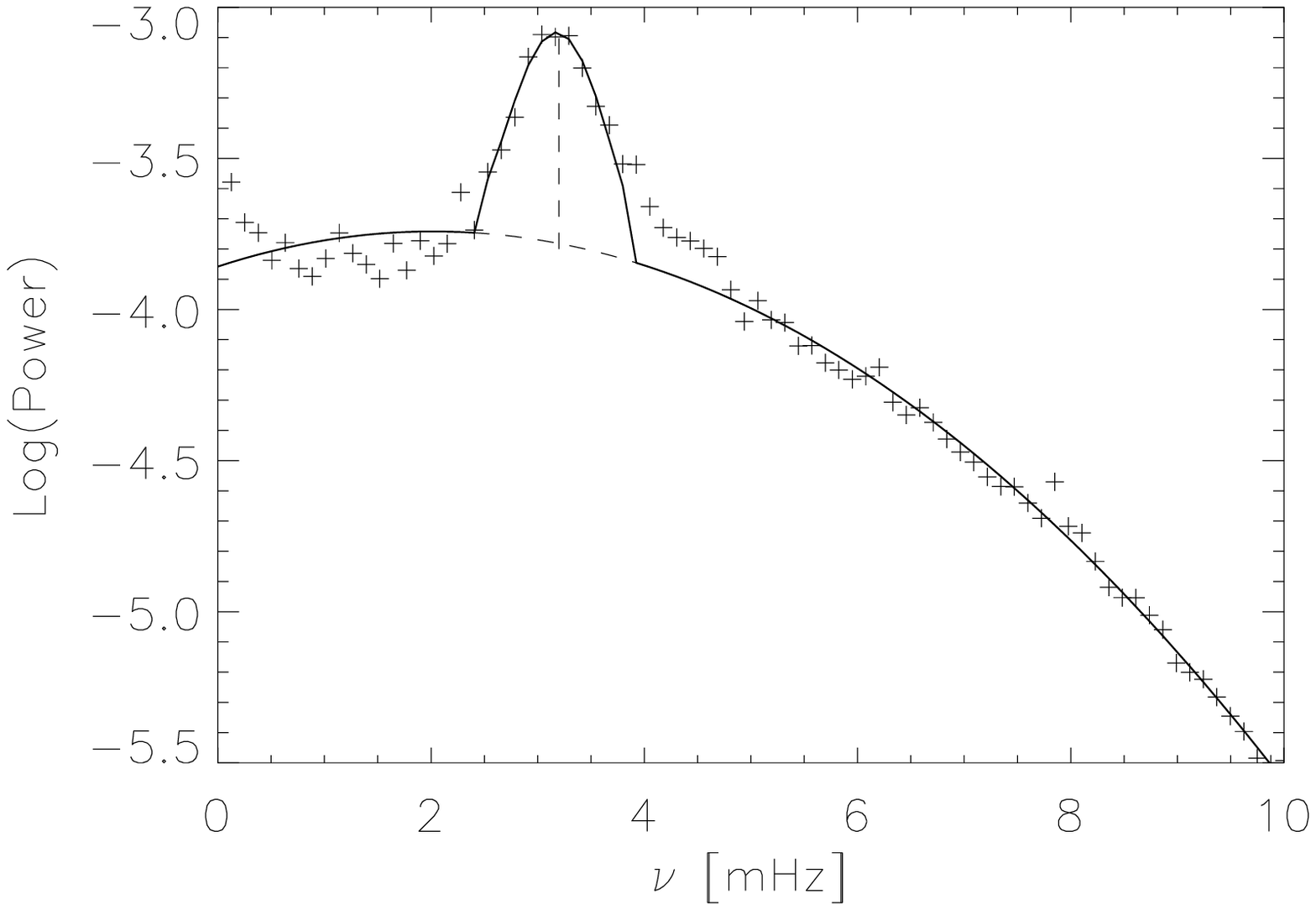}{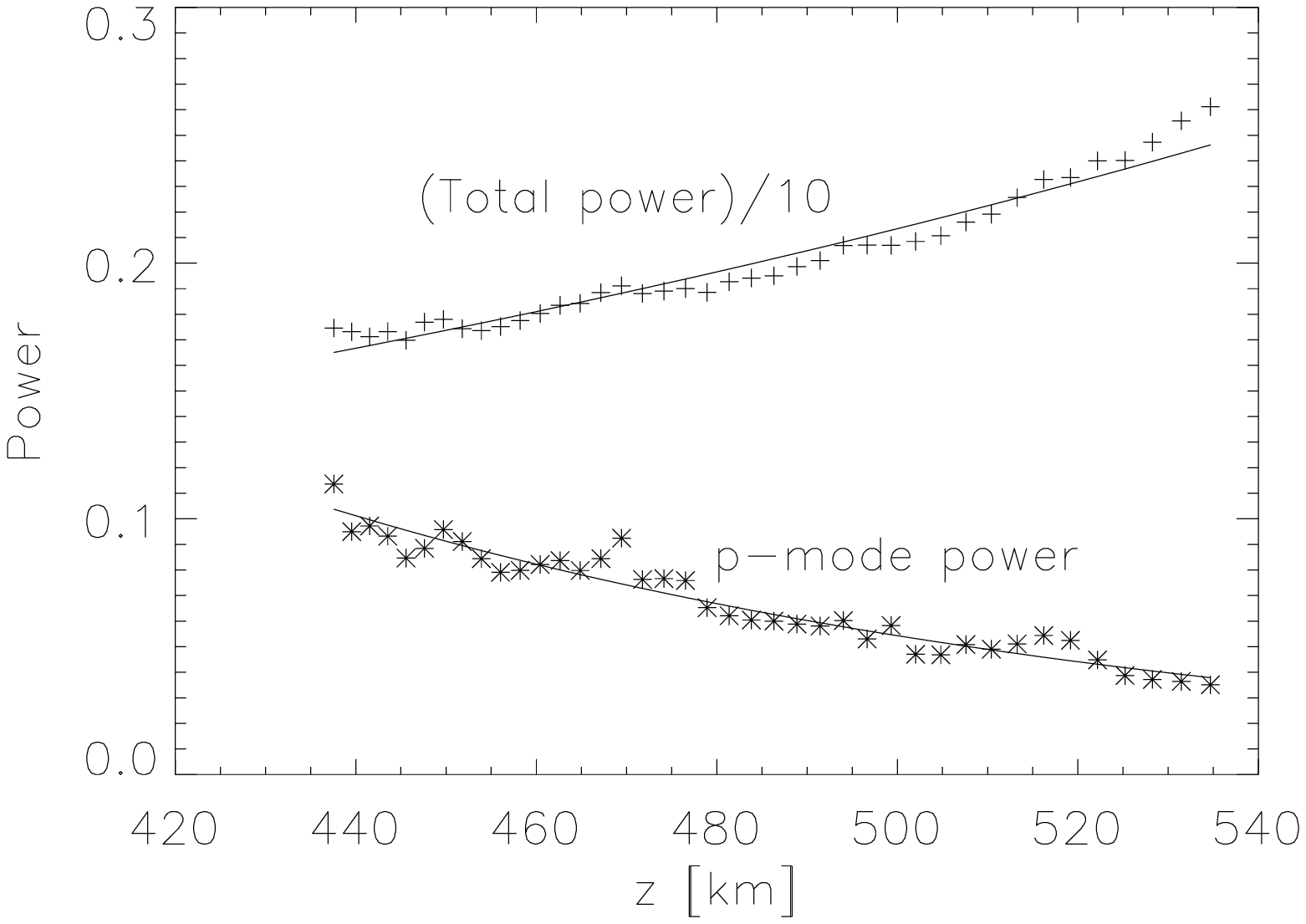}
\caption{The disk center
($\mu > 0.99$)
CO Doppler shift shows a strong peak
in the
power spectrum (left) at
$\nu=3.2$~mHz,
consistent with the solar p-mode frequencies;
the symbols show the measured power.
The background power is fit with a simple
polynomial and drops smoothly with increasing frequency,
and the p-mode peak can be roughly fit with a Gaussian function.
No evidence is seen for 3-minute period oscillation power near
$\nu=5$~mHz.
The figure on the right shows how the total power
(background plus p-mode peak)
and the power in the p-mode peak
behave at different heights in the solar atmosphere.
The power spectra fit parameters from different
disk positions were corrected for 
projection and smearing effects,
and then the
$\phi_{I-V}$
values were used to determine a physical height
corresponding to a particular value for 
$\mu$
(see text for details).
The total power increases with height
with a scale height of
$H=230$~km,
and the p-mode peak power drops with height,
consistent with an evanescent wave
and with a scale height of only
$H=90$~km,
}
\end{figure}

\clearpage
\begin{figure}
\plotone{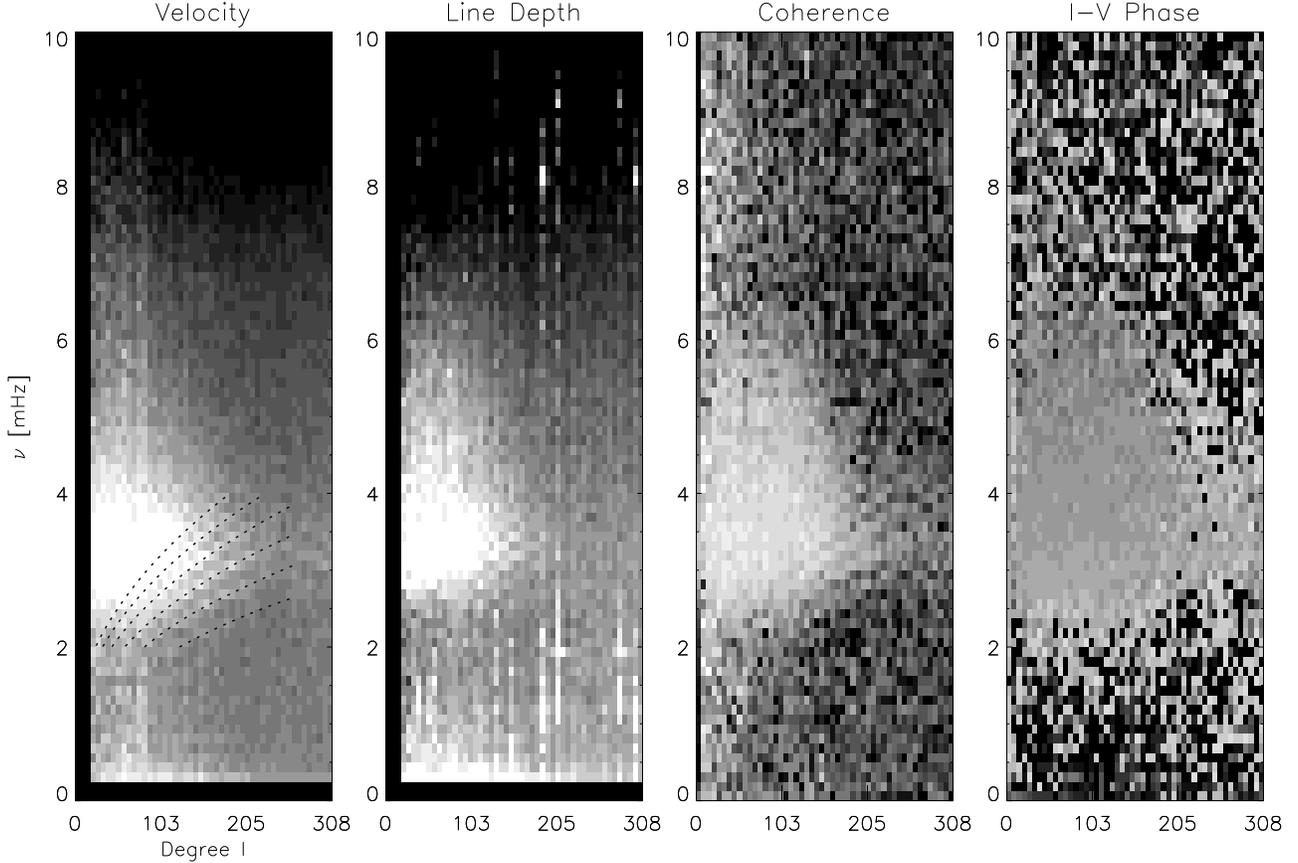}
\caption{Diagnostic 
l-$\nu$
diagrams
for various parameters in this study,
taken from a region near the center of the solar disk.
From left to right we present
the CO velocity diagram,
the CO line depth diagram,
the coherence between the line depth and velocity,
and the phase between the line depth and velocity.
The global p-mode ridges
from measurements of the NSO GONG project
for n=2 through n=7 are displayed
velocity diagram,
and exactly correspond to power ridges seen there.
Neither the velocity nor the line depth diagrams show significant
power near
$\nu=5$~mHz,
although the coherence diagram has a significant value above the noise
up to about
$\nu=6$~mHz,
and the phase shows signal levels above the noise up to 
these higher frequencies as well.
The values for the phase start near 
$\phi_{I-V}=130$~degrees
at low frequencies
$\nu=2.5$~mHz,
and drop to values of about
$\phi_{I-V}=90$~degrees
at the higher frequencies.
$\nu=4.5$~mHz,
}
\end{figure}

\clearpage
\begin{figure}
\plottwo{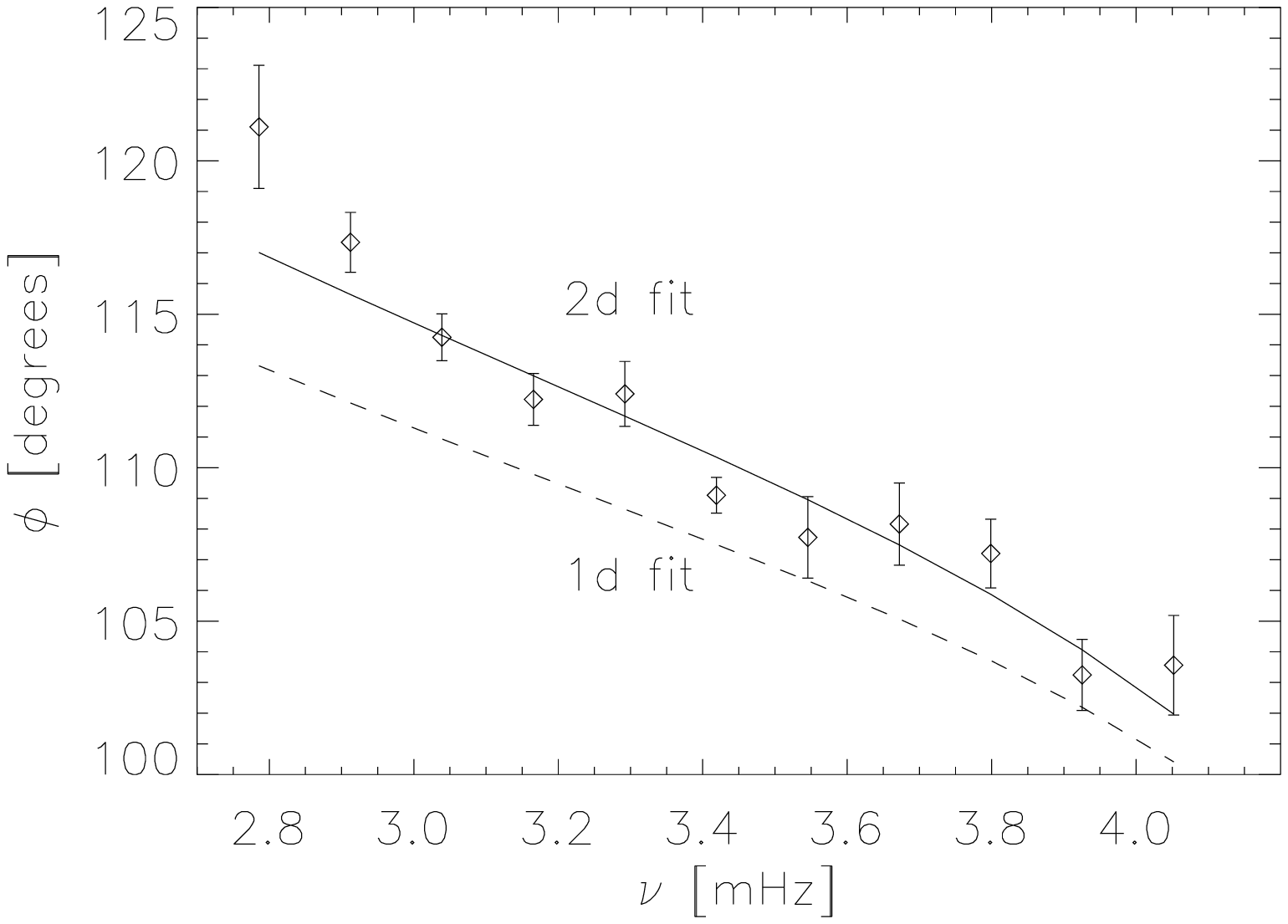}{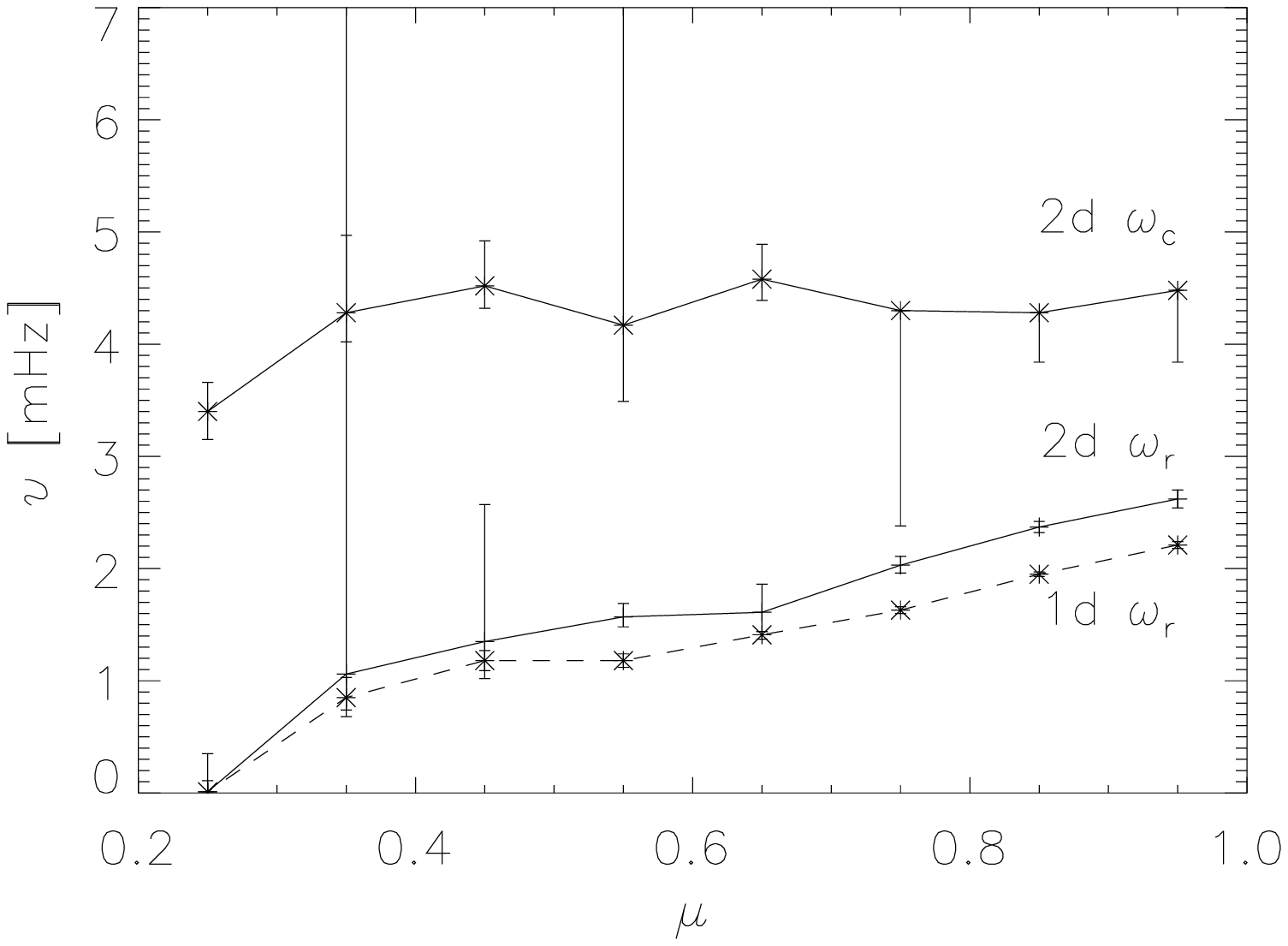}
\caption{Fits to the phase spectra are shown in the
figure on the left,
and the changes of the fit parameters across the solar disk
are plotted in figure on the right.
The phase spectra
$\phi_{I-V}(\nu)$
were fit with an expression developed
in the Appendix in two ways:
a "1-dimensional" fit was done with a constant
acoustic cutoff frequency,
$\omega_c / (2\pi) =4.5$~mHz
and finding the minimum 
$\chi^2$
value to determine
the thermal relaxation frequency,
$\omega_r$.
This fit is shown as the dashed line,
and has a large value of 
$\chi^2$.
The behavior of the fit parameter
$\omega_r / (2\pi)$
across the solar disk from these "1d" fits is shown as the dashed
line in the figure on the right.
A "2-dimensional" fit was done using a
$\chi^2$
surface to fit both the acoustic and relaxation frequencies.
This provides a better fit to the phase spectrum on the left, 
and the fit parameters are shown as they vary across the solar disk
(and presumably with height in the solar atmosphere)
on the right.
In the right figure the error bars are 68\% confidence levels determined 
from the variations of 
$\chi^2$.
}
\end{figure}


\begin{thebibliography}{}

\bibitem[Al, Bendlin \& Kneer(1998)]{al1998}
Al, N., Bendlin, C. \& Kneer, F., 1998,
\aap, 336, 743

\bibitem[Ascensio Ramos et al.(2003)]{ascensio2003}
Ascensio Ramos, A., Trujillo Bueno, J., Carlsson, M. \& Cernicharo, J. 2003,
\apj, 588, L61

\bibitem[Ayres \& Rabin(1996)]{ayres1996}
Ayres, T.R. \& Rabin, D.S. 1996,
\apj, 460, 1042

\bibitem[Cram, Brown \& Beckers(1977)]{cram1977}
Cram, L.E., Brown, D.R. \& Beckers, J.M. 1977,
\aap, 57, 211

\bibitem[Deubner et al.(1990)]{deubner1990}
Deubner, F.L., Fleck, B., Marmolino, C. \& Severino, G. 1990,
\aap, 236, 509

\bibitem[Deubner et al.(1992)]{deubner1992}
Deubner, F.L., Fleck, B., Schmitz, F. \& Strauss, Th. 1992,
\aap, 266, 560

\bibitem[Jones et al.(2002)]{jones2002}
Jones, H.P., Harvey, J.W., Henney, C.J., Hill, F., Keller, C.U. 2002,
ESA SP-505, 15

\bibitem[Kopp(1990)]{kopp1990}
Kopp, G.1990,
Ph.D. thesis, Stanford Univ.

\bibitem[Kopp et al.(1992)]{kopp1992}
Kopp, G., Lindsey, C., Roellig, T.L., 
Werner, M.W., Becklin, E.E. \& Orrall, F.Q. 1992,
\apj, 203, 203 

\bibitem[Krijger et al.(2001)]{krijger2001}
Krijger, J.M., Rutten, R.J., Lites, B.W., Strauss, Th., Shine, R.A. \& Tarbell, T.D. 2001,
\aap, 379, 1052

\bibitem[Lites \& Chipman(1979)]{lites1979}
Lites, B.W. \& Chipman, E.G. 1979,
\apj, 231, 570

\bibitem[Lites, Rutten \& Kalkofen(1993)]{lites1993}
Lites, B.W., Rutten, R.J. \& Kalkofen, W. 1993,
\apj, 414, 345

\bibitem[Massiello, Marmolino \& Strauss(1998)]{massiello1998}
Masiello, G., Marmolino, C. \& Straus, Th. 1998,
ESA SP-418, 261

\bibitem[Noyes \& Hall(1972)]{noyes1972}
Noyes, R.W. \& Hall, D.N.B. 1972,
\apj, 176, L89

\bibitem[Oliviero et al.(1999)]{oliviero1999}
Oliviero, M., Severino, G., Strauss, Th., Jefferies, S.M. \& Appourchaux, T. 1999,
\apj, 516, L45

\bibitem[Pierce(1964)]{pierce1964}
Pierce, A.K. 1964,
Applied Optics, v3, n12, 1337

\bibitem[Press et al.(1994)]{press1994}
Press, W.H., Teukolsky, S.A., Vetterling, W.T. \& Flannery, B.P. 1994,
Numerical Recipes in C, Cambridge University Press, Cambridge

\bibitem[Ruiz Cobo, Rodr\'{i}guez Hidalgo \& Collados(1997)]{ruizcobo1997}
Ruiz Cobo, B., Rodr\'{i}guez Hidalgo I. \& Collados, M. 1997,
\apj, 488, 462

\bibitem[Schmeider(1976)]{schmeider1976}
Schmeider, B. 1976,
Sol. Phys., 47, 435

\bibitem[Simoniello et al.(2008)]{simoniello2008}
Simoniello, R., Jim\'{e}nez-Reyes, S.J., Garc\'{i}a, R.A. \& Pall\'{e}, P.L. 2008,
AN, 329, No.5, 494

\bibitem[Solanki et al.(1996)]{solan1996}
Solanki, S.K., Livingston, W., Muglach, K. \& Wallace, L. 1996,
\aap, 315, 303

\bibitem[Strauss et al.(1999)]{strauss1999}
Strauss, Th., Severino, G., Deubner, F.-L., Fleck, B., Jefferies, S.M. \& Tarbell, T. 1999,
\apj, 516, 939

\bibitem[Tian et al.(2010)]{tian2010}
Tian, H., Potts, H.E., Marsch, E., Attie, R. \& He, J-S. 2010,
\aap, 519, A58

\bibitem[Uitenbroek(2000a)]{uiten2000a}
Uitenbroek, H. 2000,
\apj, 531, 571

\bibitem[Uitenbroek(2000b)]{uiten2000b}
Uitenbroek, H. 2000,
\apj, 536, 481

\bibitem[Uitenbroek, Noyes \& Rabin(1994)]{uiten1994}
Uitenbroek, H., Noyes, R.W. \& Rabin, D.S. 1994,
\apj, 432, L67

\bibitem[Vernazza, Avrett \& Loeser(1981)]{vernazza1981}
Vernazza, J.E., Avrett, E.H. \& Loeser, R. 1981,
\apjs, 45, 63

\bibitem[Wallace \& Livingston(2003)]{wallace2003}
Wallace, L. \& Livingston, W. 2003,
NSO Technical Report 03-001, NSO Tucson

\bibitem[Wedemeyer-B\"{o}hm et al.(2005)]{wedemeyer2005}
Wedemeyer-B\"{o}hm, S., Kamp, I., Bruls, J. \& Freytag, B. 2005, 
\aap, 438, 1048

\bibitem[Wiedemann et al.(1994)]{wiedemann1994}
Wiedemann, G.R., Ayres, T.R., Jennings, D.E. \& Saar, S.H. 1994,
\apj, 423, 806

\bibitem[Worrall(2002)]{worrall2002}
Worral, G. 2002,
\mnras, 335, 628



\end{thebibliography}
\end{document}